\documentclass[prl,twocolumn,showpacs,superscriptaddress,amsmath]{revtex4}
\usepackage{amsfonts,amssymb,pifont}
\usepackage{times,mathptm}
\usepackage[dvips]{graphicx}

  \usepackage{color}

\begin{document}

\title{Surmounting Barriers: The Benefit of Hydrodynamic Interactions}

\author{Christoph Lutz}
\affiliation{%
2.\ Physikalisches Institut, Universit\"at Stuttgart,
Pfaffenwaldring 57, D-70550 Stuttgart, Germany
}
\author{Michael Reichert}
\affiliation{%
Fachbereich Physik, Universit\"at Konstanz,
D-78457 Konstanz, Germany
}
\author{Holger Stark}
\affiliation{%
Fachbereich Physik, Universit\"at Konstanz,
D-78457 Konstanz, Germany
}
\author{Clemens Bechinger}
\affiliation{%
2.\ Physikalisches Institut, Universit\"at Stuttgart,
Pfaffenwaldring 57, D-70550 Stuttgart, Germany
}

\date{\today}

\begin{abstract}
We experimentally and theoretically investigate the collective
behavior of three colloidal particles that are driven by a
constant force along a toroidal trap. Due to hydrodynamic
interactions, a characteristic limit cycle is observed. When we
additionally apply a periodic sawtooth potential, we find a novel
caterpillar-like motional sequence
that is dominated by hydrodynamic interactions
and promotes the surmounting of potential barriers by the particles.
\end{abstract}


\maketitle

Hydrodynamic interactions (HI) play an important role whenever two
or more particles move 
in a viscous fluid \cite{Happel1973,Dhont1996}. 
Due to their long-range nature, they
govern the dynamics of colloidal suspensions, e.g., during self-
and collective diffusion \cite{diffusion}, sedimentation
\cite{sedimentation}, and aggregation processes \cite{Tanaka2000}.
Furthermore, HI can lead to pattern formation of rotating motors\
\cite{Grzybowski2000} with a possible two-dimensional melting
transition\ \cite{Lenz2003} and they are indispensable for the
locomotion of microorganisms \cite{Purcell1977,helices} or in the transport
of fluid by beating cilia\ \cite{cilia}.
While in all these examples many colloids are
involved, the effect of HI in few-particle
systems has been investigated only recently. It
has been demonstrated that HI mediate the correlated motion
of a pair of colloids trapped in optical tweezers\ \cite{twoparticles}
and that they give rise to interesting collective behavior,
e.g., periodic or almost periodic motions in time \cite{periodic} or even
transient chaotic dynamics in sedimenting three-particle clusters\
\cite{Janosi1997}.


In this Letter, we experimentally and theoretically demonstrate how HI
lead to a novel motional behavior of a colloidal
system comprised of at most three
particles. In contrast to the aforementioned examples, where the
colloids exhibit {\it either} deterministic drift {\it or} Brownian
diffusion, in the following we concentrate on a non-equilibrium system
where both deterministic {\it and} stochastic motions are of importance.
This work is partially motivated by a recent theoretical analysis of particles
driven by a constant tangential force along a toroidal trap
\cite{Reichert2004}. Owing to HI, the particles first go through a
transient regime and then enter a characteristic limit cycle. Here, we
present the first experimental confirmation of these findings. 
Our main objective, however, is to investigate experimentally and 
theoretically 
how the collective motion of interacting
particles changes when a sawtooth potential is
added to the constant driving force. Sawtooth potentials are an
important component for thermal ratchets studied, e.g., in connection with
biological motors\ \cite{biomotors}. Here, we demonstrate that, due to HI,
two-particle clusters exhibit an unexpected caterpillar-like motion
which facilitates the surmounting of potential barriers. This motional
sequence is largely dominated by hydrodynamic interactions in the system.

Tangential driving forces were exerted on colloidal particles with
a single three-dimensional laser tweezer that scans a circle inside our 
sample cell with the help of computer-controlled mirrors at a 
frequency $f_{\text{T}}$.
In contrast to high scanning speeds ($f_{\text{T}}>200~\text{Hz}$),
where the particle motion is entirely diffusive \cite{Lutz2004}, it
has been experimentally and theoretically demonstrated that, at
intermediate scanning speeds 
where the particle cannot directly
follow the trap, a small periodic force is transferred
from the passing optical trap onto the particle \cite{Faucheux1995}. As a
result, the particle moves with a mean velocity
${v_0}\propto f_{\text{T}}^{-1}$. For the experimental parameters in
our setup ($\lambda=532~\text{nm}$, $P\approx 200~\text{mW}$,
$f_{\text{T}}=76~\text{Hz}$), this yields ${v_0}\approx
7~\mu\text{m/s}$ for a single silica sphere with radius
$a=1.5~\mu\text{m}$ immersed in ethanol solution. Since the
particle displacement by a single kick from the optical trap is
estimated to be only about $0.08a$, the particle motion is rather
smooth. In addition, the focus size slightly changes along the circle
which in total leads to a variation of $v_{0}$ smaller than $20\%$.
This means that we can also view
$v_0$ as the result of a constant driving force $k_{0} = 6\pi \eta a v_{0}$.
To avoid wall effects which
further complicate the theoretical treatment of HI, the focus of
the laser beam was about $40~\mu\text{m}$ above the bottom plate
of our sample cell. The particles were illuminated with a white
light source and imaged to a CCD camera, which was connected to a
PC where images were compressed and stored. Particle trajectories
were obtained off\-line by a particle tracking algorithm\ 
\cite{videomicroscopy}. To enhance the electrostatic
coupling between the charged spheres, no salt was added to the suspension.

\begin{figure}
\includegraphics[width=0.7\columnwidth]{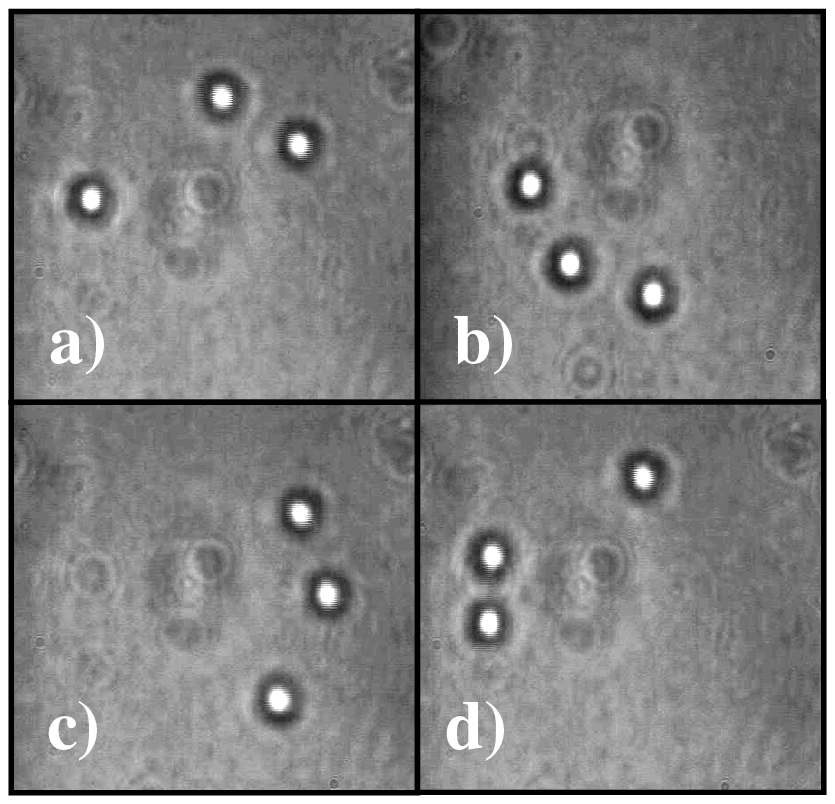}
\caption{Snapshots of a video sequence \cite{videohome}
describing the characteristic limit cycle of three colloidal particles
(bright) driven along a toroidal trap in counter-clockwise
direction. The time difference between the pictures is 4.8~s each. The
optical trap is blocked with optical filters.}
\label{fig:snapshots}
\end{figure}

A sequence of typical snapshots ($\Delta t=4.8~\text{s}$) of three
particles in the periodic limit cycle driven counter-clockwise on
a circular ring with $9.86~\mu\text{m}$ radius is shown in
Fig.~\ref{fig:snapshots}. The two-particle cluster at the top
[Fig.~\ref{fig:snapshots}(a)]
catches up with a preceding third particle
[Fig.~\ref{fig:snapshots}(b)] until they form a triplet for a
short time. Since the middle particle in this triplet is most
efficiently screened from the fluid flow, it pushes the frontmost
particle ahead so that the two front particles leave the last one
behind [Fig.~\ref{fig:snapshots}(c)]. Due to drag reduction by
drafting, this cluster is then again catching up with the single
particle [Fig.~\ref{fig:snapshots}(d)]. We also investigated the
transition from the unstable particle configuration, where the
particles had originally the same distance, into the periodic
limit cycle and found very good agreement with theoretical
predictions (data not shown) \cite{Reichert2004}.
It should be also mentioned that the limit cycle described above is 
entirely absent when repeating the experiment close to a surface. We
believe that this is a result of the rigid boundary that alters 
HI close to surfaces \cite{Diamant2005}.

\begin{figure}
\includegraphics[width=0.92\columnwidth]{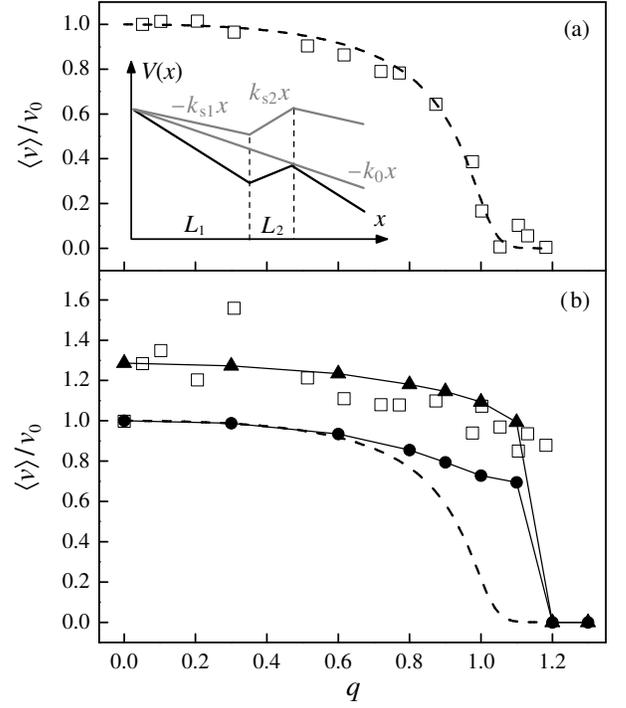}
\caption{(a) Normalized single-particle velocities as obtained from
experiments ($\square$) and from the analytical expression (\ref{eq:v})
(-\hspace{2pt}-\hspace{2pt}-). Inset: Schematic representation of the
tilted sawtooth potential. (b) Corresponding two-particle-cluster
velocities from experiments ($\square$) and numerical simulations
including HI (\ding{115}) and without HI (\ding{108}).
For comparison, the dashed curve of (a) is replotted.}
\label{fig:velocity}
\end{figure}

After having demonstrated hydrodynamic effects in the presence of
a constant driving force, we now want to study the cooperative particle
motion in the presence of a more complicated force profile. 
In addition to the constant driving force
$k_{0}>0$, we apply a sawtooth potential $V_{\text{s}}(x)$ with
period $L$, where $x$ is the arc-length coordinate along the
circumference of the trap.
In the first segment of length $L_{1}$,
$V_{\text{s}}(x)$ exerts an additional force $k_{\text{s}1} > 0$
on the particle, thus enhancing the drift motion, whereas in the
second segment of length $L_{2}=L-L_{1}$, the force
$-k_{\text{s}2}<0$ counteracts $k_{0}$. As a result, the particle
is moving in a tilted sawtooth potential $V(x)$, as illustrated in
the inset of Fig.~\ref{fig:velocity}(a). The single-particle
motion is described by the one-dimensional Smoluchowski equation 
from which the probability distribution
$p(x,t)$ for finding the particle at position $x$ in the tilted
sawtooth potential is calculated.
Since the particle moves in a circle, $p(x,t)$
is stationary. After solving the Smoluchowski equation on the two
segments using appropriate matching conditions at $x=0,L$ and
$x=L_{1}$, we calculate the constant probability current
$j$ and finally arrive at the mean particle
velocity $\langle v\rangle = j L$\ \cite{extended}:
\begin{eqnarray}
\frac{\langle v\rangle}{v_{0}} & = & \left[1 +
\frac{q^{2}}{(1-q)[1-(1-q) \delta]} \right. \nonumber \\
 & & \qquad \left. \times \left(\delta - \frac{1}{C}
\frac{1-e^{-(1-q)\delta C}}{(1-q)[1-(1-q) \delta]} \right) \right]^{-1} .
\label{eq:v}
\end{eqnarray}
Here, $v_{0}=k_{0}/(6\pi\eta a)$ is the particle velocity without
sawtooth potential, and $\delta=L_{2}/L$ describes the asymmetry
of the potential. $C=k_{0}L/(k_{\text{B}}T)$ is the energy
dissipated on the length $L$ relative to the thermal energy
$k_{\text{B}}T$. It is linked to the conventional Peclet number
$\text{Pe}=Ca/L$. Since $C\gg 1 $ in our
case, one might naively expect a purely deterministic motion. In
general, however, the type of motion depends on the value
$q=k_{\text{s}2}/k_{0}$ which serves as a measure for the
amplitude of the sawtooth potential. For $q<1$, the motion is
indeed purely deterministic and determined by the first term in
the second line of Eq.~(\ref{eq:v}). At $q=1$, the net force acting on
the particle in segment $L_2$ vanishes ($k_{0}-k_{\text{s}2}=0$), and
the colloid moves entirely stochastically until it resumes the drift motion
in segment $L_1$. At $q>1$, it even experiences a potential
barrier that gives rise to a stick-slip motion. Fig.~\ref{fig:velocity}(a) 
(dashed line) shows the result of
Eq.~(\ref{eq:v}) as a function of $q$ for the set of experimental
parameters specified below. Note that, in the limit
$C\rightarrow\infty$, the Boltzmann tail at $q>1$ vanishes and
$\langle v\rangle$ becomes zero at $q=1$.

\begin{figure}
\includegraphics[width=0.95\columnwidth]{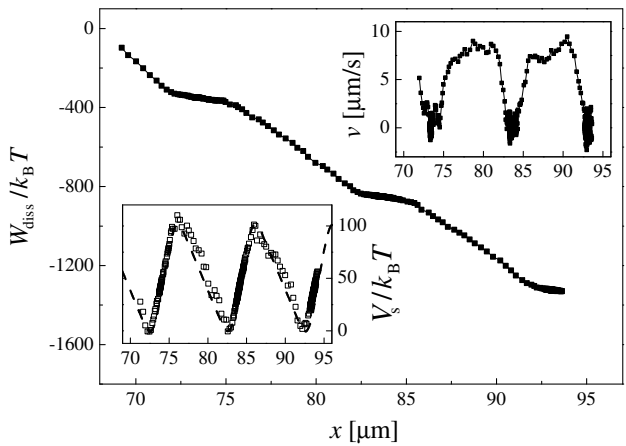}
\caption{Experimentally determined dissipated energy of a particle
in a sawtooth potential with a constant driving force. Upper
inset: position-dependent velocity of a single particle. Lower
inset: sawtooth potential as obtained from the dissipated energy
($\square$) and as calculated from the intensity variation along the
contour of the toroidal trap (-\hspace{2pt}-\hspace{2pt}-).}
\label{fig:potential}
\end{figure}

To realize the situation described above experimentally, we weakly
modulated ($\leq \pm 12\%$) the intensity of the scanning optical tweezer. 
This was achieved with an electro-optical-modulator
controlled by a function generator that was synchronized with the
scanning motion of the laser focus. Before discussing hydrodynamic
coupling of particles in such a situation, let us briefly
demonstrate that our experimental approach allows to simultaneously 
apply a constant drift force and a quasi-static periodic potential to 
the colloids. The upper inset of Fig.~\ref{fig:potential} shows the
position dependent velocity $v(x)$ of a single particle (averaged
over $200~\text{ms}$ each) determined from its trajectory. From
this, we calculate the position dependent force
$k(x)=6\pi\eta a v(x)$ and obtain the energy dissipated by the
particle, $W_{\text{diss}}(x)=-\int_0^x k(x')dx'$
(Fig.~\ref{fig:potential}), which is essentially identical to the
tilted sawtooth potential $V(x)$\ \cite{footnote}. 
After subtraction of the linear 
contribution $k_{0}x$ corresponding to the mean particle velocity 
$v_0$ measured without intensity modulation, we finally obtain the 
spatially periodic potential acting on the particles. The open symbols 
of the lower inset in Fig.~\ref{fig:potential} show the corresponding sawtooth
potential. On the other hand,
taking into account the size of the laser focus and
that of the particles (both effects lead to some rounding of the
edges in the potential), we can calculate from the sawtooth-shaped
modulation of the laser intensity the effective potential acting
on the particles\ \cite{extended}. It is shown as dashed line and 
demonstrates the good agreement with our measurements. The amplitude of 
the potential corresponds to an intensity modulation amplitude of $\Delta
P\approx 20~\text{mW}$ that is consistent with the measured intensity 
variation.

First, we experimentally determined the normalized mean particle
velocity $\langle v\rangle/v_{0}$ of a single particle [see Eq.~(\ref{eq:v})] 
as a function of the amplitude of the sawtooth potential, the
latter being proportional to $q$ (with the proportionality
constant as fitting parameter). The results are plotted in
Fig.~\ref{fig:velocity}(a) as open symbols and were obtained for
the following experimental parameters: $v_0=7.24~\mu\text{m/s}$
and $\delta={L_2/L}\approx 2.7~\mu\text{m}/10.3~\mu\text{m}$ which
yields $k_{0}=245\cdot 10^{-15}~\text{N}$ and $C\approx 610$.
The dashed curve in Fig.~\ref{fig:velocity}(a) is the result of a
least-mean-square fit to Eq.~(\ref{eq:v}) that shows excellent
agreement with the data when $L_2$ as a free fitting parameter
assumes the value $L_2=1~\mu\text{m}$ and thus $\delta=0.1$.
We attribute the deviation from the experimental $L_{2}$ to the
difference between the experimentally realized and the perfect sawtooth 
potential as mentioned above.

\begin{figure}
\includegraphics[clip=true,angle=-90,width=0.95\columnwidth]{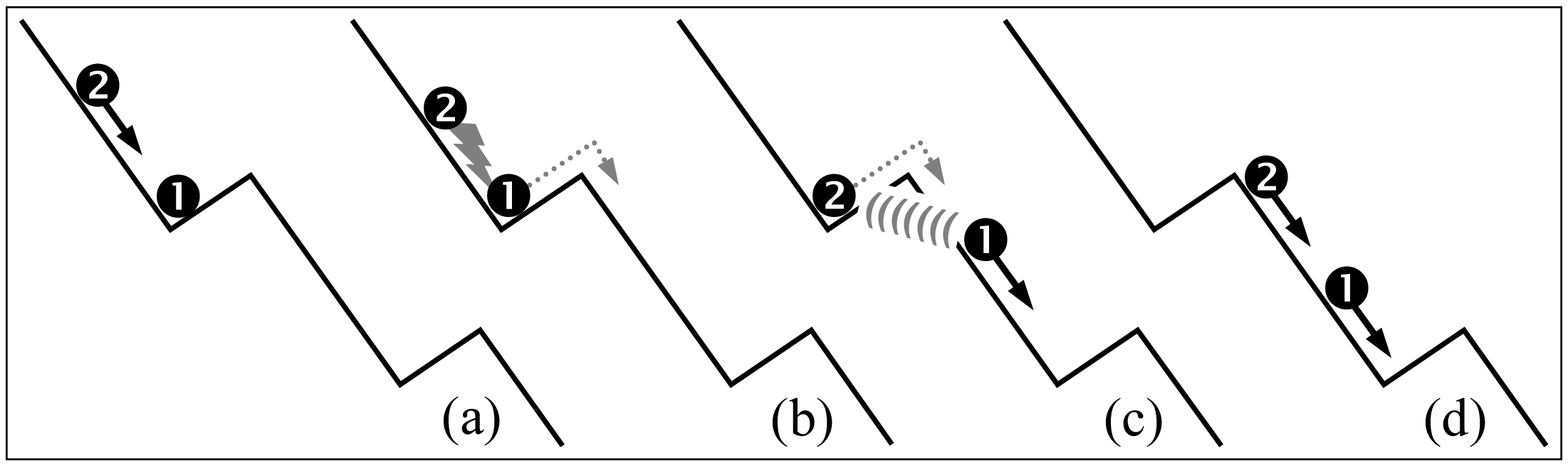}
\caption{Sketch of the motional sequence of a cluster comprised of
two particles. Due to HI and electrostatic interactions, a
caterpillar-like motion is observed which facilitates the
surmounting of potential barriers.} \label{fig:caterpillar}
\end{figure}

Next, we investigated the case where three particles were
driven along the sawtooth potential. Similar to the case
$V_{\text{s}}(x)\equiv 0$, the particles change their relative
distances as a function of time (for video sequences see\ 
\cite{videohome}). In contrast to
Fig.~\ref{fig:snapshots}, however, we do not observe the same periodic
limit cycle but a novel type of collective motion induced by
the sawtooth landscape. Figure~\ref{fig:velocity}(b) shows the average 
velocity of two-particle clusters in a tilted sawtooth potential as
determined from experiments (open symbols) and from Brownian-dynamics
simulations (solid triangles) with the above parameters. 
In our image analysis, we defined a cluster as a pair of
particles with center-to-center distance $\leq 5a$, 
corresponding to arc-length distance $\leq 0.7L$.
Obviously, the two-particle velocity is larger than the corresponding 
one-particle velocity which is again a result of the reduced
hydrodynamic friction. What is actually surprising, is the fact that 
even when the single-particle velocity drops to zero around $q\ge 1$
(because the particle becomes trapped in a potential well), the
cluster velocity varies only by about 20\% in that $q$-range.
This clearly demonstrates that the surmounting of potential barriers
is largely facilitated for particle clusters compared to a single
colloid. The reason for this behavior is due to HI that
lead to an interesting, caterpillar-like motion of the clusters as
described in the following. Assume particle ``2''
[Fig.~\ref{fig:caterpillar}(a)] drifts into a potential well which
is already occupied by particle ``1''. Due to a combined effect of
electrostatic repulsion and HI, particle ``2'' will push particle
``1'' over the barrier [Fig.~\ref{fig:caterpillar}(b)]. This in turn
causes a hydrodynamic drag that particle ``1'' exerts onto particle
``2'' [(Fig.~\ref{fig:caterpillar}(c)] which 
pulls particle ``2'' across the barrier
[Fig.~\ref{fig:caterpillar}(d)]. The motional sequence just
described is reminiscent to that of a caterpillar which first
stretches out, adheres at the front, and then pulls the tail
towards the head. Once such a mode is initiated (e.g., by a
thermally induced process), the outlined motion may last over
several periods $L$ until the particles become trapped in separate
potential minima due to thermal fluctuations of their distance.
Then, the motion stops until the situation
described in Fig.~\ref{fig:caterpillar}(a) is initiated by a
spontaneous hop of particle ``2'' into the potential well occupied by 
particle ``1''.

This motional pattern of two-particle clusters was also observed
in our Brownian-dynamics simulations. The electrostatic
inter-particle repulsion is described by a screened Coulomb
potential \cite{DLVO} $V_{\text{rep}}(r)
=[Ze^{\kappa a}/(1+ \kappa a)]^{2}
\lambda_{\text{B}}k_{\text{B}}Te^{-\kappa r}/r$
with Debye length $\kappa^{-1} \approx 300~\text{nm}$,
where the effective particle charge is adjusted to $Z\approx 8000$ 
(taking $\lambda_{\text{B}}\approx 0.7~\text{nm}$ for the
Bjerrum length in water) so that the center-to-center distance between
the particles does not go below about $3a$ as observed in the
experiment. The Brownian-dynamics simulations including HI 
were carried out using a predictor-corrector-type integration 
scheme which is first order in the time step $\Delta t$\ \cite{extended}.
For particle separations $r\ge 3a$, the mobilities describing the HI and thus
the mutual coupling of all particles  can be well approximated by the
Rotne-Prager tensor \cite{Dhont1996}, which is the far-field
expansion up to order $1/r^{3}$ (for details, see Ref.~\cite{extended}).

To demonstrate the crucial role of HI for the enhanced cluster motion,
we performed numerical simulations without HI. 
The mean particle velocity, defined as the average over
{\it all} particles {\it and} times, coincides with the single-particle
curve in Fig.\ \ref{fig:velocity} when plotted as a function of $q$. 
This is clear since, due to the average over all particles, the electrostatic 
forces cancel each other (actio = reactio) and thus, on average, the particles 
move independently. 
The mean velocities of the particle {\it clusters}, however, deviate from
the single-particle velocities, as illustrated in Fig.~\ref{fig:velocity}(b) 
(solid circles).
Nevertheless, as expected, the
average cluster velocities are always below the corresponding curve
where HI are included.
For $q\ge 1$, the two-particle-cluster motion without HI consists
of singular events where particle ``2'' moves down the seceding flank
of the tilted sawtooth potential and pushes particle ``1'' over the
potential barrier [Fig.~\ref{fig:caterpillar}(a,b)].  In contrast to
the caterpillar-like behavior, the collective motion stops here
since the drag due to HI is missing. Then, another thermal activation 
is necessary to induce the sequence again\ \cite{videohome}.

Finally, we point out that the short-range electrostatic
repulsion between the particles is only of minor importance for 
the collective, caterpillar-like motion as described above.
It is rather the long-ranged HI that dominate the force which pushes
the front particle in Fig.~\ref{fig:caterpillar}(b) over the barrier
(note that HI decay asymptotically as $1/r$ whereas the dominant term in 
the repulsive electrostatic force is exponentially damped, 
$F_{\text{rep}}(r) = -\nabla_{r} V_{\text{rep}}(r)
\propto e^{-r/\lambda_{\text{D}}}/r$ with $\lambda_{\text{D}}\ll a$).
The caterpillar-like motion was also obtained in simulations where
the strength of the electrostatic interaction potential was reduced by a 
factor of 50 so that the minimum distance to contact of the particles was
$0.1a$\ \cite{extended}.

The present study demonstrates that HI strongly dominate the motional
behavior of driven colloidal particles. In the presence of a constant
driving force, we experimentally confirmed the recently predicted
characteristic limit cycle. If in addition a static periodic potential
is applied along the toroidal trap, we find that colloidal clusters
perform a caterpillar-like motion which facilitates the surmounting of
large potential barriers. This novel type of motion which is
predominantly triggered by HI is also confirmed by numerical simulations.
In addition to a better microscopic understanding of HI in few-particle
systems, 
our results are also of relevance for the motion of interacting particles 
in thermal ratchets, which are considered as models for biological motors.
Indeed, recent in-vivo experiments demonstrate
that the speed of coupled motor proteins is increased compared to the speed 
of a single motor\ \cite{Kural05}.

\acknowledgments{
We would like to thank V.\ Blickle for helpful discussions.
This work was supported by the Deutsche Forschungsgemeinschaft through
SFB Transregio 6 and Grant No.\ Sta 352/5-1.
}


\begin{thebibliography}{99}
\bibitem{footnote}
$W_{\text{diss}}$ also contains small contributions from the stochastic
motion of the particles that can be neglected relative to its absolute
value.
\end{thebibliography}
\end{document}